\\

Title: Gamma-analysis of airborne particulates sampled in Youzhno-Sakhalinsk town at March – April 2011

Authors: E. G. Tertyshnik, V. P. Martynenko, F. A. Andreev, G. B. Artemyev

The experience of discovery of the radioactive products which have released into atmosphere of Sakhalin region from Fukushima Daiichi accident is presented. Sampling of airborne particulates and atmosphere fallout was carried out by means of the air ventilation set and horizontal gauze planchs, respectively. The HPGe detector was used for gamma analyses of the airborne samples. Since 23 March we confidently measured $^{131}$I in the airborne samples, after 03.04.2011 we also registered a rise of activity $^{137}$Cs and $^{134}$Cs. $^{132}$Te and $^{132}$I were discovered in ashen sample of the planch, which had exposed in Youzhno-Kurilk from 14 to 17 March. The effect of the pairs production when in the samples $^{208}$Tl presence, which emits gamma-quanta of 2615 keV, causes a rise in apparatus spectra of the peak corresponding to energy 1593 keV, which could be in error ascribed to $^{140}$La. It had been experimentally shown that the systematic reduction of $^{134}$Cs content in measuring samples due to effect of gamma - gamma coincidence did not exceed 7 % (for the detector and geometry of the measurement used).

-------------------

Заголовок: Гамма-спектрометрический анализ проб аэрозолей, отобранных в Южно-Сахалинске в марте-апреле 2011 г.

Авторы: Э.Г. Тертышник, В.П. Мартыненко, Ф.А. Андреев, Г.Б. Артемьев

Представлен практический опыт обнаружения радиоактивных продуктов, поступивших в атмосферу Сахалинской обл. при аварии на АЭС Фукусима Дайичи. Отбор проб аэрозолей и атмосферных выпадений осуществляли с помощью вентиляционной фильтрующей установки и горизонтальных марлевых планшетов, соответственно. Гамма-спектрометрический анализ отобранных проб выполняли, используя HPGe- детектор. Начиная с 23марта, в пробах аэрозолей надёжно фиксировалось наличие $^{131}$I, после 05.04.2011г. наблюдалось также повышенное содержание $^{137}$Cs и $^{134}$Cs . При анализе пробы планшета, который экспонировался в Южно-Курильске с 14 по17 марта, было отмечено присутствие $^{132}$Te и $^{132}$I. Эффект образования пар при наличии в пробах $^{208}$Tl, испускающего гамма-кванты с энергией 2615 кэВ, приводит к появлению в аппаратурном спектре пика с энергией 1593 кэВ, который может быть ошибочно приписан $^{140}$La. Экспериментально показано, что систематическое занижение содержания $^{134}$Cs в пробах аэрозолей и атмосферных выпадений из-за эффекта гамма-гамма совпадений не превышало 7 % (для использованного детектора и геометрии измерения).

\\

**Гамма-спектрометрический анализ проб аэрозолей, отобранных в Южно-Сахалинске в марте-апреле 2011 г.**

Э.Г. Тертышник, В.П. Мартыненко, Ф.А. Андреев, Г.Б. Артемьев (ФГБУ «НПО «Тайфун»)

В результате аварии на японской АЭС Фукусима Дайичи (Fukushima Daiichi, 37.4214$^0$ с.ш., 141.0325$^0$ в.д.), которая произошла 11 марта 2011 г. в результате катастрофического землетрясения и цунами, в атмосферу поступили радионуклиды — продукты деления и активации. В связи с этим потребовались объективные данные о уровнях радиоактивного загрязнении воздуха для информирования населения и администрации соседнего с Японией о. Сахалин.

Поскольку на территории Сахалинской области не было гамма-спектрометров высокого энергетического разрешения с полупроводниковыми германиевыми детекторами, гарантирующими надёжную информацию о содержании гамма-излучающих радионуклидов в пробах со сложным изотопным составом, руководство Федеральной службы по гидрометеорологии и мониторингу окружающей среды и ФГБУ «НПО «Тайфун» решило направить на о. Сахалин автомобильную лабораторию радиационной разведки (АЛРР), предназначенную для метеорологического и радиационного мониторинга в районе радиационно-опасного объекта и оборудованную гамма-спектрометром с HPGe- детектором [1]. АЛРР вместе с обслуживающими её сотрудниками была доставлена самолётом МЧС России в Южно-Сахалинск 20.03.2011.

Гамма-спектрометр в составе АЛРР скомпонован на основе аппаратуры фирмы ОРТЕК США (EG & G ORTEC) включающей: многоканальный анализатор импульсов DigiDART с цифровой обработкой сигнала, обеспечивающей наивысшую пропускную способность и энергетическое разрешение при высоких загрузках, число каналов анализатора 16384, использовали в работе 8192 канала; детектор из особо чистого германия марки GEM-2070 с энергетическим разрешением 1,8 кэВ по линии 1332 кэВ $^{60}$Co и относительной эффективностью 26 % по той же линии.

Для защиты детектора от внешнего гамма-излучения детектора применяли свинцовый экран с толщиной стенок 50 мм. С помощью программы спектрометрического анализа MAESTRO-32, осуществляли поиск и идентификацию фотопиков, а также расчет их интенсивности.

Калибровку спектрометра по фотоэффективности регистрации гамма-квантов различных энергий проводили с помощью объёмного источника типа ОИСН (эталонный Стандартный источник 2-го разряда), изготовленного в виде гранул, которые можно помещать в измерительные контейнеры, чтобы моделировать геометрии измерений по форме и размерам соответствующим измеряемым пробам аэрозолей. Изготовление и аттестация калибровочных источников выполнена во Всероссийском научно-исследовательском институте физико-технических и радиотехнических измерений (ВНИИФТРИ). Относительная расширенная неопределённость удельной активности Стандартного источника составляла 6 % при доверительной вероятности 95 %.

Ежедневно, начиная с 21 марта 2011 г., сотрудники ФГБУ «НПО «Тайфун» выполняли гамма-анализ проб атмосферных выпадений и проб аэрозолей, отобранных в Южно-Сахалинске (46.95° с.ш., 142.717° в.д.) в течение суток, а также проб выпадений из Южно-Курильска (44.0306° с.ш., 145.856° в.д.). Для отбора проб аэрозолей использовали стационарную воздухо-фильтрующую установку (ВФУ) Сахалинского Управления по Гидрометеорологии и Мониторингу Окружающей Среды (Сахалинское УГМС). За сутки с посредством этой ВФУ выделяли аэрозоли из объёма приземного воздуха порядка 20000 м$^3$, аэрозоли осаждались на фильтре из фильтроткани марки ФПП-15-1,5. Пробы атмосферных выпадений, отбирали с помощью горизонтальных марлевых планшетов, рабочая площадь планшетов 0,3 м$^2$ [2]. Снятые после экспонирования фильтры и марля планшетов не озоляли, чтобы исключить потери радиойода при повышенной температуре, а формировали в брикеты диаметром 46 мм и толщиной 8 мм путём прессования с применением соответствующей пресс-формы и ручного гидравлического пресса.

Выброшенные при аварии на Фукусима Дайичи $^{132}$Te и $^{132}$I , были обнаружены 21.03.2011 г. в атмосферных выпадениях при анализе золы планшета из Южно-Курильска, удалённого от аварийной АЭС на расстояние примерно 840 км. Этот планшет был установлен 14.03. и снят 17.03.2011 г., в золе планшета была зафиксирована аномально высокая бета-активность и золу направили на гамма-анализ. Активность $^{132}$I на момент измерения составляла 0,56 Бк. Период полураспада этого радионуклида всего 2,295 часа, а анализ проводили спустя 5 суток, поэтому $^{132}$I регистрировался как дочерний продукт распада $^{132}$Te (период полураспада 3,204 суток). В пробе золы $^{131}$I не обнаружен (период полураспада 8, 04 сут.). После удаления йода в результате воздействия высокой температуры при озолении планшета радиоактивное равновесие в паре $^{132}$Te — $^{132}$I восстанавливается через 12 часов и активность $^{132}$I становиться равной активности

$^{132}$Te. (равновесное отношение активностей $^{132}$I/$^{132}$Te = 1,03). Полагая, что при озолении планшетов теллур полностью сохранился, плотность выпадений $^{132}$Te на 17.03.2011 в Южно-Курильске составила примерно 8 Бк/м$^2$. При расчётах эффективность улавливания марлевых планшетов принята равной 0,7.

Фрагменты аппаратурного спектра этой пробы приводятся на рис. 1, хорошо виден пик, соответствующий $^{132}$Te (энергия 228,3 кэВ), а также пики, обусловленные $^{132}$I (667,7 и 522,7 кэВ). Таким образом, нами было зарегистрировано первое поступления продуктов деления на территорию России от аварийных реакторов на АЭС Фукусима Дайичи, которое имело место в период 14 — 17.03.2011 г. Над территорией США газообразный $^{133}$Xe ($T_{1/2}$ = 5,25 сут.) был обнаружен 16.03.2011 г. [3], время переноса газообразных выбросов на расстояние 7700 км составило 4 —5 суток. Первое поступление, связанных с аэрозолями $^{131}$I, $^{132}$Te, $^{132}$I, $^{137}$Cs, $^{134}$Cs, зарегистрировано в США 17 — 18.03.2011 г. [4].

Наши результаты измерения плотности атмосферных выпадений в Южно-Курильске и Южно-Сахалинске, а также результаты измерения объёмной активности воздуха опубликованы в работе [5].

Начиная с 23.03.2011 г. на аэрозольных фильтрах, через которые фильтровался воздух с помощью ВФУ, установленной в Южно-Сахалинске, уверенно регистрировался $^{131}$I. Динамика изменения содержания $^{131}$I, а также $^{137}$Cs и $^{134}$Cs в марте – апреле 2011г. представлена на рис. 2 и 3. Поместив последовательно за аэрозольным фильтром марки ФПП-15-1,5 сорбционно-фильтрующий материал, предназначений для улавливания молекулярного йода, марки СФМ-И, установили, что доля молекулярной фракции $^{131}$I в нижнем слое атмосферы Южно-Сахалинска составляла примерно 20 % суммарного содержания (молекулярная + аэрозольная составляющие). Максимальная зафиксированная нами объёмная активность $^{131}$I во вдыхаемом воздухе в Южно-Сахалинске (870 мкБк/м$^3$) в 8 тыс. раз меньше допустимой среднегодовой объёмной активности этого радионуклида в воздухе для населения (для $^{131}$I $ДОА_{нас}$ = 7,3 Бк/м$^3$, а для $^{134}$Cs $ДОА_{нас}$ = 19 Бк/м$^3$ ) [6].

Возрастание содержания $^{131}$I в период 24 — 29 марта не сопровождалось ростом содержания $^{137}$Cs и $^{134}$Cs, по-видимому, в это время температура ядерного топлива была ниже температуры интенсивного испарения цезия, а резкое увеличение содержания $^{131}$I в атмосфере Южно-Сахалинска после 05.04.2011 г. сопровождалось соответствующим ростом содержания $^{137}$Cs и $^{134}$Cs (рис. 3), что может свидетельствовать о существенном увеличении температуры в аварийных реакторах в это время. В пробе аэрозолей, отобранной 07 – 08.04.2011 г., обнаружен $^{136}Cs$ ($T_{1/2}$ = 13,16 сут.), содержание которого составило 12 ± 6 мкБк/м$^3$. Используя высокочувствительные детекторы (большого объёма) с хорошим экранированием и анализирую объединённую за месяц пробу аэрозолей, авторы работы [7] также зарегистрировали в атмосфере Италии индикаторные количества $^{136}Cs$ (4,6 ± 1,4 нБк/м$^3$ ).

В процессе отбора проб аэрозолей с помощью ВФУ на аэрозольных фильтрах накапливаются дочерние продукты распада $^{222}Rn$ (радона) и $^{220}Rn$ (торона) и гамма-излучение этих продуктов могут создавать помехи при анализе проб аэрозолей. Практически во всех аппаратурных спектрах, полученных при анализе проб аэрозолей, присутствуют хорошо выраженные пики, которые соответствуют энергиям 583 и 2614 кэВ, что однозначно указывает о наличии в пробах $^{208}Tl$. Гамма-кванты с энергией 2614,5 кэВ испускаются при распаде $^{208}Tl$ с почти 100 % вероятностью и при взаимодействии с материалом детектора образуют пары (электрон и позитрон). Образование пар может происходить в поле ядра или электрона [8], образование пар в поле ядра иногда называют когерентным (упругим), а если часть энергии воспринимается каким-нибудь электроном атома, имеет место ионизация атома (неупругое взаимодействие) и в спектре может проявиться пик, обусловленный вылетом за пределы детектора квантов характеристического излучения германия. На рис. 4 приведен фрагмент аппаратурного спектра, зарегистрированного при анализе пробы аэрозолей 25.03.2011 г., экспозиция 11100 с. Здесь же представлен участок аппаратурного спектра от препарата $^{232}Th$, на обоих спектрах виден пик двойного вылета аннигиляционных квантов (энергия 1593 кэВ). Этот пик может быть ошибочно приписан $^{140}La$ (1596,5 кэВ), присутствие которого ожидается в результате ядерных аварий [9]. Пик, соответствующий энергии 1588 кэВ, обусловлен гамма-квантами $^{228}Ac$ (квантовый выход 0,032 %), а пик соответствующий энергии 1583 кэВ, вероятно, вызван присутствием $^{214}Bi$ (квантовый выход 0,007 %). В табл. 1 приведены отношения площади пика двойного вылета аннигиляционных квантов к площади пика, соответствующего энергии 2614,5 кэВ для двух типов детекторов. Поскольку увеличение объёма детектора приводит к возрастанию эффективности регистрации в пике полного поглощения (2614,5 кэВ), а 95 % аннигиляционных квантов вылетают за пределы детектора только из верхнего сравнительно тонкого слоя германиевого кристалла (~ 3 мм), отношение площади пика двойного вылета к площади пика, соответствующего энергии 2614,5 кэВ уменьшается при увеличении эффективного объёма детектора. Данные табл. 1 могут использоваться для оценки вклада пика вылета в фотопик $^{140}La$.

Эффект суммирования зарядов, которые образуются в чувствительном объёме детектора при регистрации совпадающих по времени гамма-квантов $^{134}Cs$ с энергией 604,6 кэВ и 795,8 кэВ, приводит к уменьшению эффективности регистрации в соответствующих пиках полного поглощения. Если не учитывать этот эффект, возникает систематическое занижение измеряемой активности.

Таблица 1. Отношение площади фотопика, соответствующего энергии 1593 кэВ к площади пика от гамма-квантов $^{208}Tl$ (2614,5 кэВ) для измеряемого препарата диаметром 46 мм и толщиной 8 мм.

| Детектор и условия измерений | Отношение площадей фотопиков |
|---|---|
| Детектор АЛРР, GEM-2070. Измеряемая проба на торце детектора. Среднее значение для 10 проб аэрозолей | 0,13 |
| Детектор GEM-2070 без экрана. Препарат $^{232}Th$ на торце детектора. | 0,10 |
| Детектор GEM-2070 без экрана. Препарат $^{232}Th$ на расстоянии 5 см от торца детектора. | 0,10 |
| Детектор GEM-30185 экран 10 см свинца. Препарат $^{232}Th$ на торце детектора. | 0,091 |
| Детектор GEM-30185 экран 10 см свинца. Препарат $^{232}Th$ на расстоянии 5 см от торца детектора. | 0,082 |
| Детектор GEM-30185 экран 10 см свинца. Препарат $^{232}Th$ на расстоянии 10 см от торца детектора. | 0,087 |

Оценку поправки на эффект суммирования (для $^{134}Cs$) в измеренных нами пробах атмосферных выпадений и аэрозолей выполнили экспериментально. Используя раствор $^{134}Cs$, приготовили препарат, моделирующий спрессованные в брикеты фильтры (диаметр 46 мм, толщина 8 мм). Затем активность $^{134}Cs$ в препарате была определена дважды: при размещении препарата на торцевой поверхности криостата детектора ($A_0$) и при удалении препарата от детектора на расстояние 20 см ($A_{20}$). Поскольку эффект суммирования для удалённой геометрии практически отсутствует, то поправка $К_с$ (множитель) для реальных проб, измеренных в ближней геометрии (когда проба размещалась вплотную к детектору) составляет $К_с = A_{20} / A_0$.

Для детектора, размещённого в АЛРР (GEM-2070) и используемой геометрии измерений $К_с = 1,07$; следовательно, чтобы учесть эффект суммирования, результаты измерения содержания $^{134}Cs$, приведённые в [5], необходимо умножить на 1,07. Таким образом, отношение активности $^{134}Cs$ к активности $^{137}Cs$, которое составляло [5] в среднем 0,75 возрастёт после введения поправки до 0,80. Отметим, что для проб аэрозолей, отобранных на западном побережье США, отношение активности $^{134}$Cs к активности $^{137}$Cs равнялось 0,7 [4]. Эффект суммирования становится более существенным, если применяются детекторы с бóльшим чувствительным объёмом, так при использовании детектора GEM-30185 поправка для такой же геометрии составит 1,12.

**Заключение.**

1. В пробах атмосферных выпадений и пробах аэрозолей, отобранных в гг. Южно-Курильске и Южно-Сахалинске, обнаружены $^{131}$I, $^{132}$Te, $^{132}$I, $^{137}$Cs, $^{134}$Cs и $^{136}$Cs. Представлен график изменения содержания $^{131}$I, $^{137}$Cs и $^{134}$Cs (аэрозольная фракция) в период с 20 марта по 10 апреля 2011г.

2. Загрязнение атмосферного воздуха от аварии на АЭС Фукусима Дайичи в марте –апреле 2011г. не представляло опасности для населения Сахалинской обл., поскольку максимальное зафиксированное нами содержание $^{131}$I примерно в 8 тыс. раз меньше допустимой среднегодовой объёмной активности этого радионуклида в воздухе для населения.

3. Присутствие на аэрозольном фильтре заметных количеств $^{208}$Tl (характерные пики 583 и 2614 кэВ) приводит к появлению в аппаратурном спектре пика двойного вылета, которому соответствует энергия 1593 кэВ. Этот пик может быть ошибочно приписан $^{140}$La.

4. Экспериментально была определена поправка, учитывающая эффект суммирования совпадающих гамма-квантов при распаде $^{134}$Cs. Для использованной геометрии измерений (препарат диаметром 46 и толщиной 8 мм на торцевой поверхности детектора) и детектора GEM-2070 поправка равна 1,07.

Авторы выражают благодарность сотрудникам ФГБУ «НПО «Тайфун» Вакуловскому С.М. и Рогашову А.Д., а также сотрудникам Сахалинского УГМС Золотухину Е.Г. и Калиновской Л.В. за помощь в работе.

**Литература**

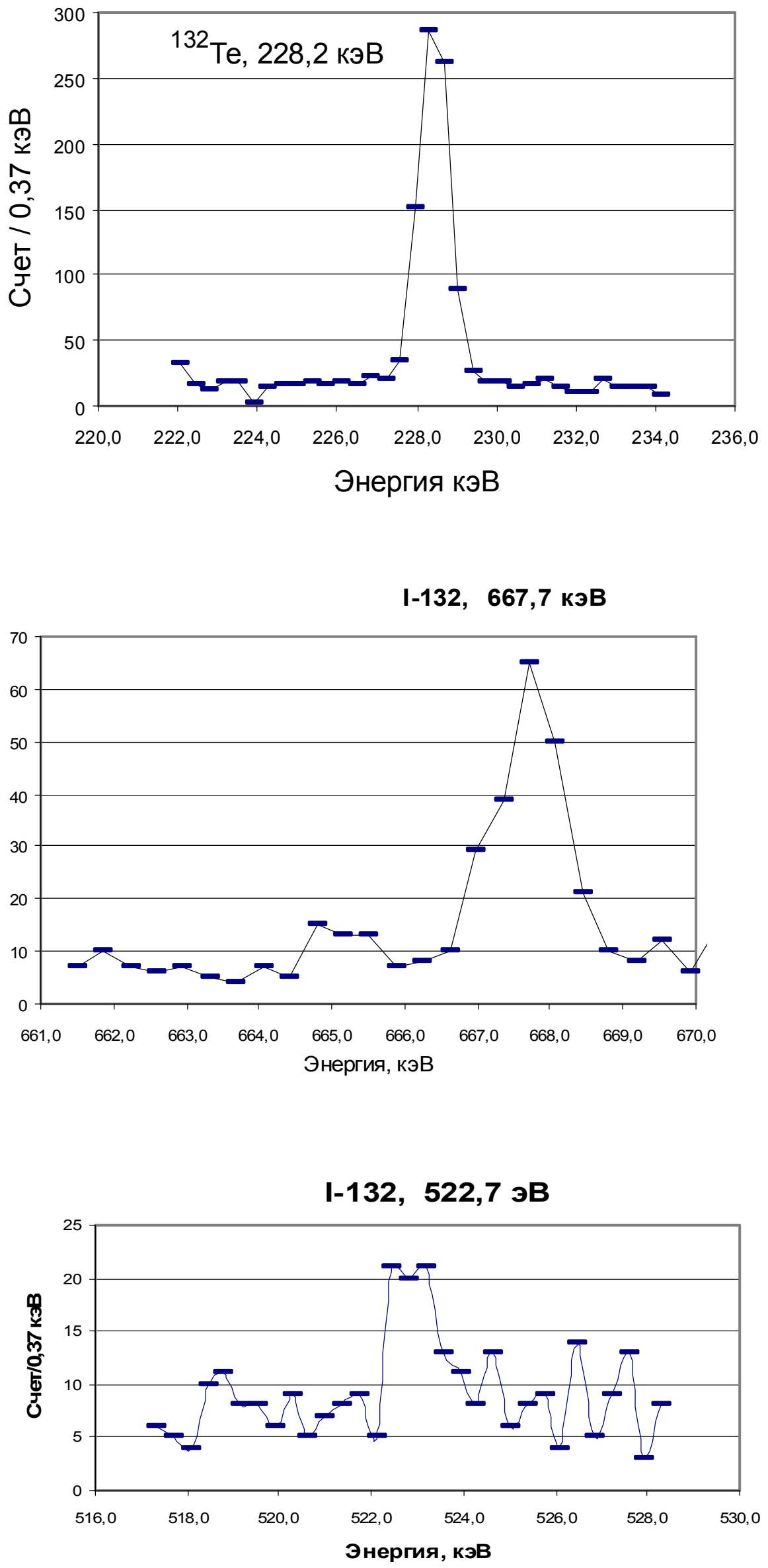

Рис. 1. Фрагменты аппаратурного спектра золы планшета, экспонированного 14 – 17 марта, время измерения 10024 с.

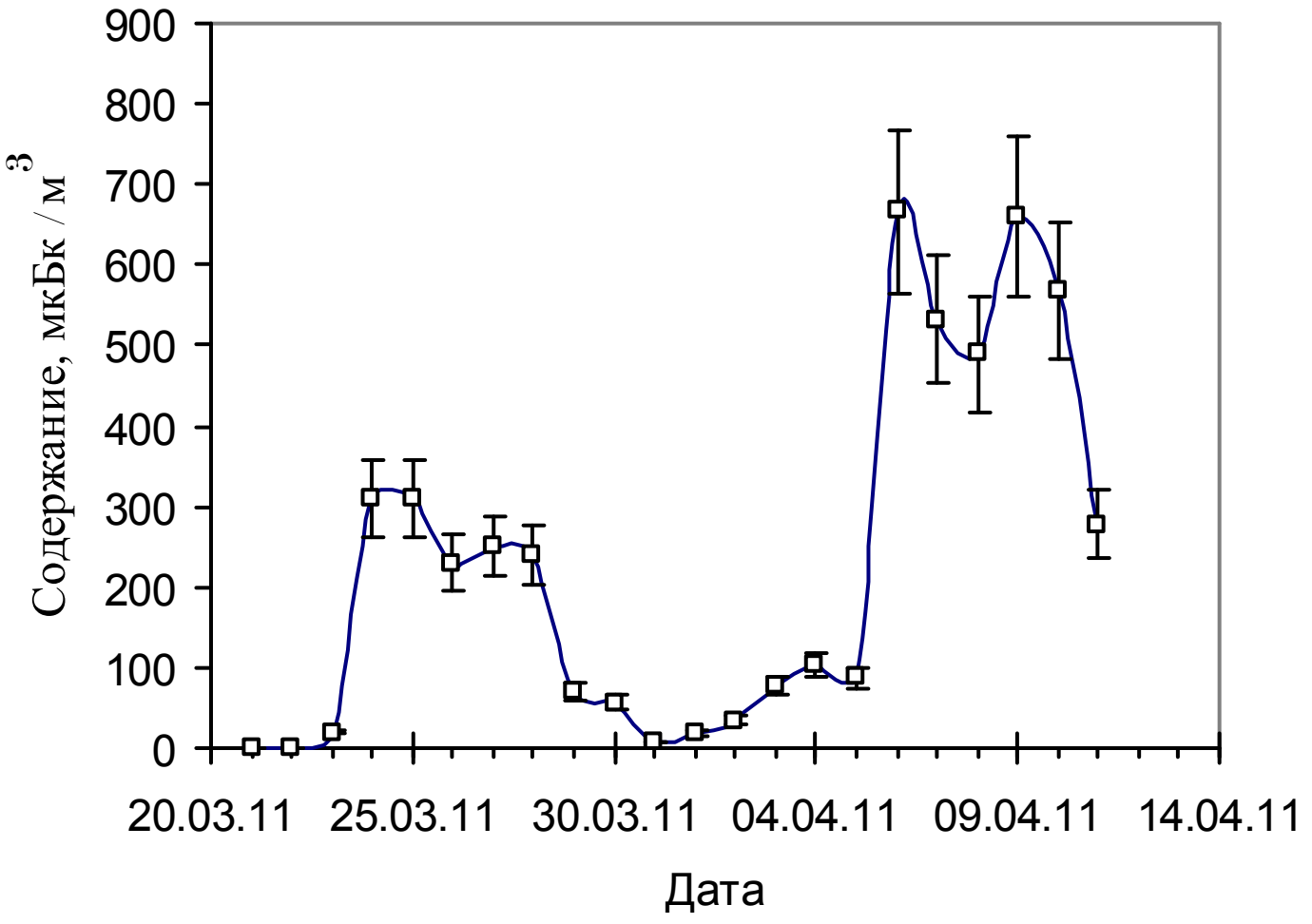

Рис. 2. Изменение со временем объёмной активности аэрозольной фракции $^{131}I$ в атмосфере Южно-Сахалинска в марте- апреле 2011 г.

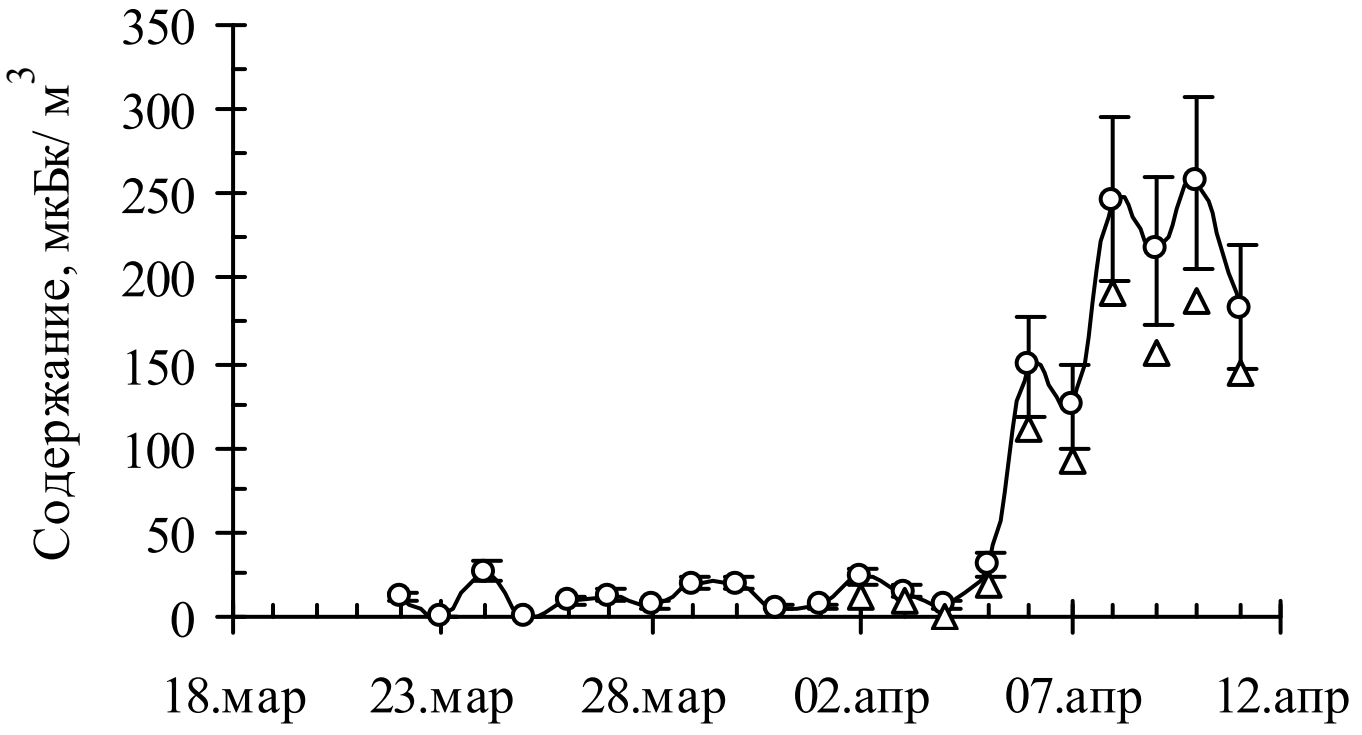

Рис. 3. Изменение содержания $^{137}Cs$ (○) и $^{134}Cs$ (Δ) в атмосфере Южно-Сахалинска со временем.

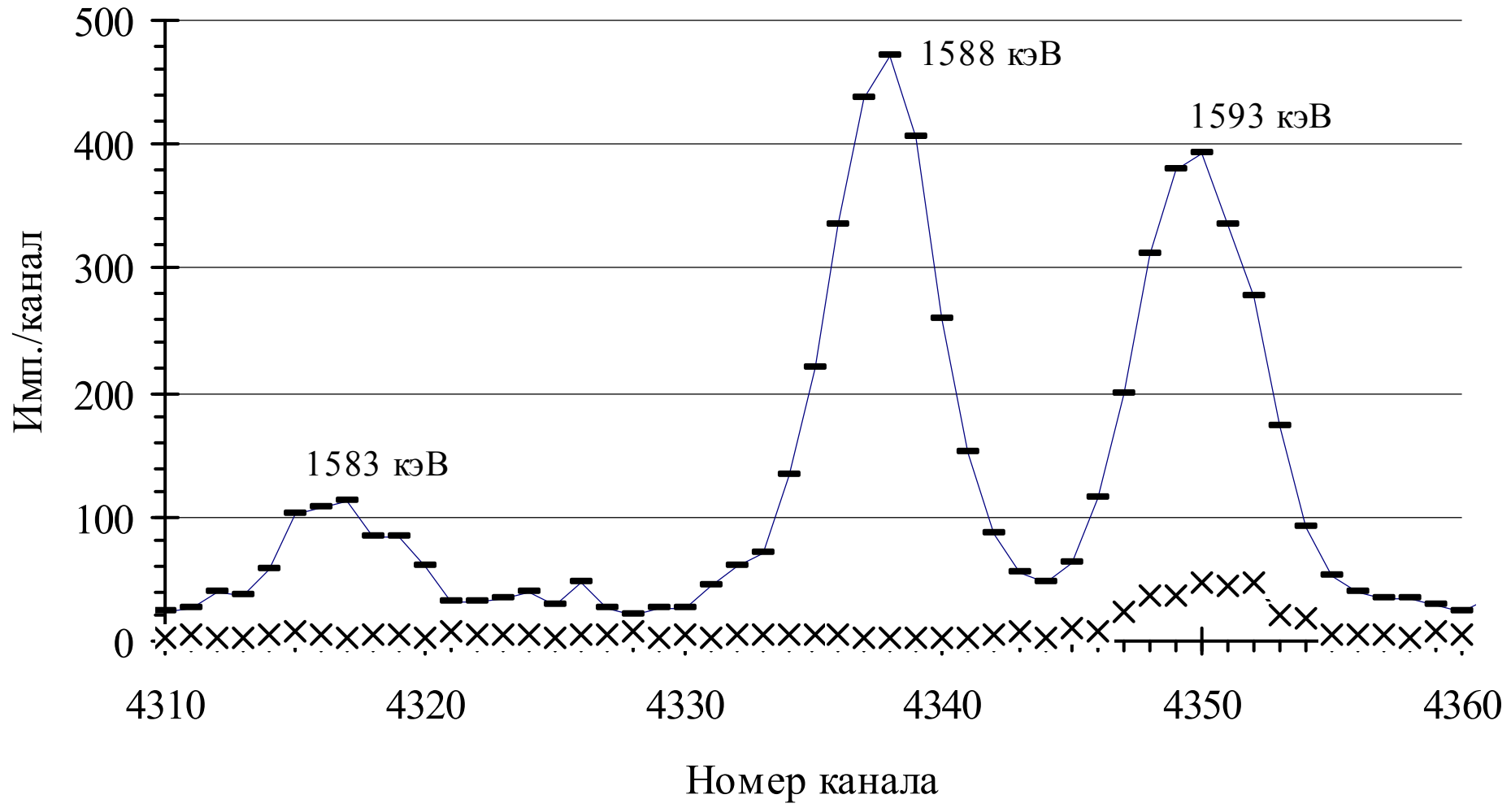

Рис.4. Фрагменты аппаратурных спектров, зарегистрированных при анализе пробы аэрозолей 25.03.2011 г. (×) и анализе препарата $^{232}$Th (—).